\documentclass[aps,prb,twocolumn,floatfix,showpacs,citeautoscript,longbibliography,hyperlinks,superscriptaddress]{revtex4-2}

\usepackage{amsmath,amssymb,amsfonts}

\usepackage{graphicx}
\usepackage{comment}

\usepackage[colorlinks=true,citecolor=blue]{hyperref}
\apptocmd{\sloppy}{\hbadness 10000\relax}{}{}
\usepackage{xcolor}

\newcommand{\revision}[1]{{{#1}}}

\begin{document}
    \title{Quantum transport in Cooper pair splitters using hierarchical equations of motion}

\author{Riya Baruah}
\affiliation{Department of Applied Physics, Aalto University, 00076 Aalto, Finland}
\author{Neill Lambert}
\affiliation{RIKEN Center for Quantum Computing, Wakoshi, Saitama 351-0198, Japan}
\author{Franco Nori}
\affiliation{RIKEN Center for Quantum Computing, Wakoshi, Saitama 351-0198, Japan}
\affiliation{Physics Department, University of Michigan, Ann Arbor, Michigan 48109-1040, USA}
\author{Christian Flindt}
\affiliation{Department of Applied Physics, Aalto University, 00076 Aalto, Finland}
\date{\today}
\begin{abstract}
We investigate charge transport in Cooper pair splitters beyond the weak-coupling and Markovian limits. To this end, we employ  hierarchical equations of motion (HEOM), which can capture the combined effects of strong coupling to the leads, nonperturbative interactions, and finite voltage and temperature differences. Within this framework, we compute the electric currents as functions of the level positions of a Cooper pair splitter for various voltage and temperature configurations. In the large-bias regime, our results reduce to analytical expressions obtained from a Markovian Lindblad equation. However, recent experiments were conducted with finite voltage or temperature differences, where a Markovian description may not suffice. In this regime, HEOM yield quantitative agreement with the measured currents. We can also account for an experimentally observed thermoelectric effect in Cooper pair splitters. Our results show that HEOM provide a useful framework for describing nonequilibrium quantum transport in Cooper pair splitters and related hybrid devices.
\end{abstract}

\maketitle

\section{Introduction}

Hybrid nanostructures, which combine superconductors with quantum dots and normal-metal leads, constitute a versatile platform for exploring correlated electron transport~\cite{DeFranceschi:2010}. Cooper pair splitters, where a superconducting contact is coupled to spatially separated normal leads via quantum dots, have attracted considerable interest as potential sources of spin-entangled electron pairs~\cite{Recher2001,Lesovik2001}. These devices distill nonlocal entanglement from Cooper pairs inside a superconductor, and they represent a promising building block for solid-state quantum information processing~\cite{Hofstetter2009,Herrmann2010,Hofstetter2011,Schindele2012,Das2012,Schindele2012,Fueloep2014,Fulop2015,Tan2015,deacon2015,Borzenets2016,Bruhat2018,Baba2018,Tan2021,Ranni2021,Pandey2021,Krtssy2022,Ranni2022,Bordoloi2022,Wang2022,Wang2023,Bordin2023,deJong2023}. Recent experiments have also demonstrated that coupled Cooper pair splitters can be used to realize short Kitaev chains~\cite{Dvir2023,tenHaaf2024,tenHaaf2025,Bordin2025} and poor-man’s Majorana bound states~\cite{Leijnse2012}, highlighting their relevance for engineered topological superconductivity~\cite{Leijnse2012b,Aguado2017}. From a theoretical perspective, Cooper pair splitters pose a challenging problem, as they combine superconductivity, strong interactions, nonperturbative couplings, and nonequilibrium transport conditions imposed by finite voltage or temperature differences.

A variety of theoretical approaches have been used to describe charge and heat transport in Cooper pair splitters. In the noninteracting or weakly interacting regime, scattering theory and the Bogoliubov–de Gennes formalism provide a transparent description of local and nonlocal Andreev processes~\cite{Blonder1982}, and they have been widely used to analyze conductance, noise, and entanglement properties~\cite{Melin2008,Chevallier2011}. More generally, nonequilibrium Green’s functions offer a systematic framework for treating coherent transport in multiterminal superconducting devices~\cite{Jacquet2020}. These approaches are particularly useful for describing phase-coherent transport, but become demanding if strong interactions must be accounted for~\cite{rech2012}.

\begin{figure}[hb!]
    \centering
\includegraphics[width=1\columnwidth]{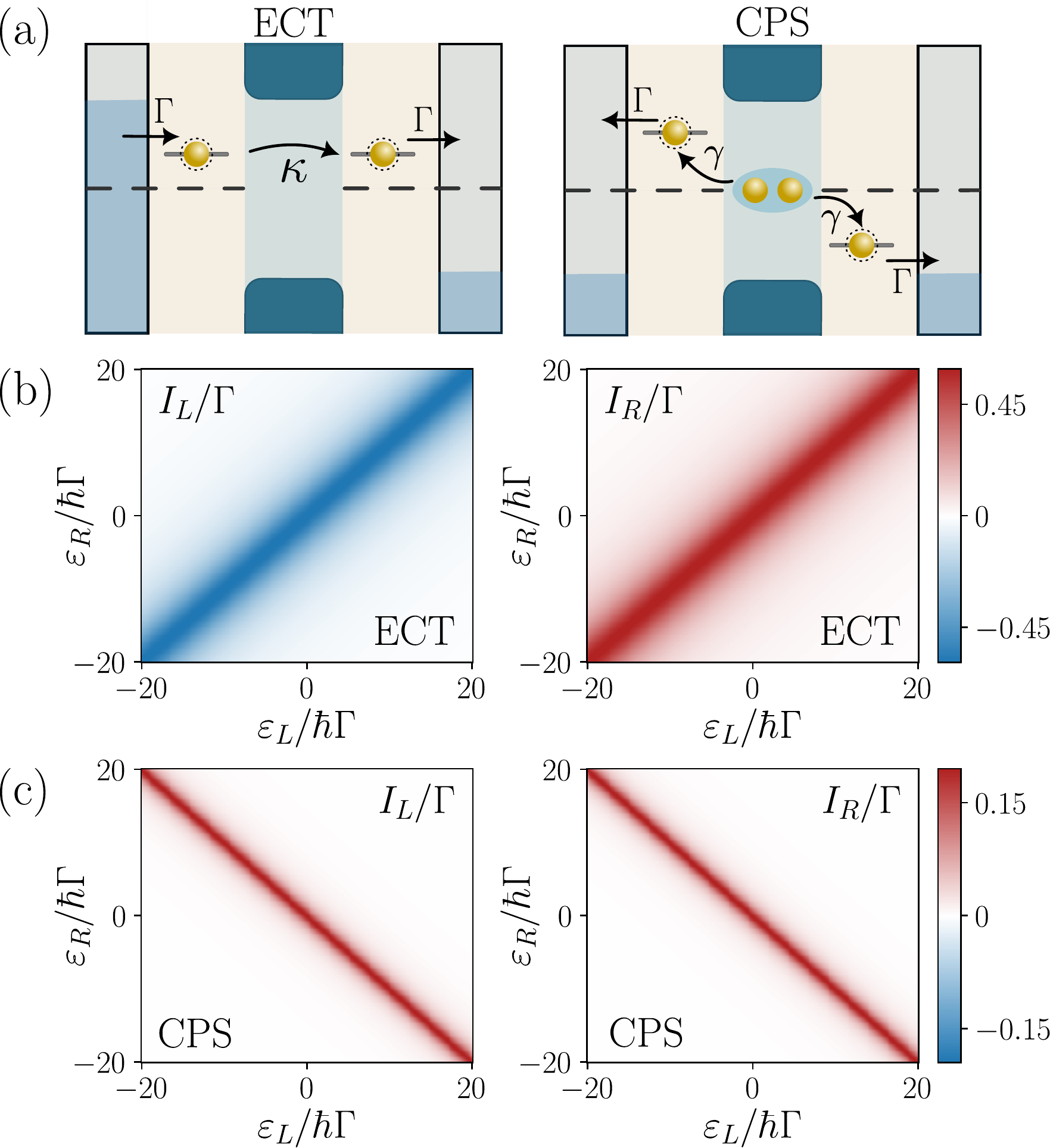}
    \caption{Quantum transport in a Cooper pair splitter. (a)~The device consists of left and  right quantum dots that are coupled to a superconductor and separate normal-state leads. If a voltage is applied across the quantum dots, and the quantum dot levels are aligned, $\varepsilon_L=\varepsilon_R$, charge transport takes place via elastic cotunneling (ECT). When the normal-state leads function as drains, charge transport takes place via Cooper pair splitting (CPS), if the quantum dot levels are asymmetrically detuned, $\varepsilon_L=-\varepsilon_R$. The amplitudes for ECT and CPS are denoted by $\kappa$ and $\gamma$, respectively, and $\Gamma$ is the rate for tunneling to the leads. (b) Particle currents in the left and right leads for ECT. (c) Particle currents in the left and right leads for CPS. The parameters are $\gamma=0.4 \hbar\Gamma$ and $\kappa=2 \hbar\Gamma$.}
\label{fig1}
\end{figure}
An alternative line of work is based on quantum master equations, which are well suited for describing interacting systems coupled to external leads~\cite{Gurvitz1996,Sauret2004,Sauret2005,Governale2008,Futterer2009,Meng2009,Eldridge2010,Hiltscher2011,Zonda2023,Hussein2016,Hussein2017,trocha2018a}. With large bias voltages, these methods can in some cases provide analytical expressions for charge currents and other transport properties, such as the noise and higher-order current correlators~\cite{Walldorf2018,Walldorf2020,Tam2021,Brange:2021,Brange:2024}. However, recent experiments on Cooper pair splitters were conducted with finite voltage or temperature differences in regimes where the coupling to the leads may not be weak. In that case, Markovian descriptions may fail to correctly describe the transport, and a more general treatment is needed.

In this article, we use hierarchical equations of motion~(HEOM) to investigate charge transport in Cooper pair splitters. Figure~\ref{fig1} shows the device geometry and examples of the electric currents for different voltage configurations. As illustrated in Fig.~\ref{fig1}(a), the Cooper pair splitter consists of two quantum dots, which are tunnel coupled to a superconductor and to left and right normal-state electrodes. Split Cooper pairs from the superconductor can tunnel into the quantum dots and further on into the normal-state electrodes, depending on the applied voltages. In Figs.~\ref{fig1}(b) and~\ref{fig1}(c), we show the currents in the leads as functions of the level positions of the quantum dots. In Fig.~\ref{fig1}(b), the device is voltage biased so that the left lead functions as a source of electrons, while the right lead serves as a drain. In that case, the electric currents can be found analytically [cf.~Eq.~(\ref{eq:1source1drain})], and they display a resonance along the diagonal, where the two quantum dot levels are aligned, and elastic cotunneling between the quantum dots is on resonance. By contrast,
in Fig.~\ref{fig1}(c), the device is biased with large voltages so that both leads function as drains for electrons on the quantum dots. The electric currents  [cf.~Eq.~(\ref{eq:2drains})] then display a resonance along the antidiagonal, where the total energy of the quantum dot levels vanishes, and Cooper pair splitting is on resonance.  

The theoretical results capture many features of the recent experiments on Cooper pair splitters~\cite{Wang2022,Wang2023,Bordin2023}. However, while the analytic expressions are obtained in the large-bias limit, the measurements were carried out with finite voltages, and the electric currents then vanish if the quantum dot levels are moved outside the transport window.  To account for the combined effects of finite voltage or temperature differences, interactions, broadening effects, and superconductivity, we here use HEOM to calculate the electric currents in the Cooper pair splitter. While HEOM have been successfully applied to a wide range of quantum impurity and transport problems~\cite{Jin2008,Zheng2012,Wang2013,Song2017,Xu2017}, their application to superconducting hybrid devices has so far remained limited. \revision{(However, Andreev molecules,
where two quantum dots are coupled to a common superconductor, have been investigated using the numerical renormalization group method~\cite{Zalom2024})}. Using HEOM, we find good agreement with the recent experiments~\cite{Wang2022,Wang2023,Bordin2023}, and we can also account for a thermoelectric electric effect observed in Cooper pair splitters~\cite{Cao2015,Burset2018,Hussein2019,Kirsanov2019,Tan2021,Arora2025,Arrachea2025}. As such, our results establish HEOM as a versatile and powerful tool for describing charge transport in Cooper pair splitters and related superconducting nanostructures.

The rest of our paper is organized as follows. In Sec.~\ref{sec:CPSmodel}, we describe our model of a Cooper pair splitter consisting of a superconductor coupled to separate normal-metal leads via quantum dots. In Sec.~\ref{sec:fif}, we introduce the fermionic influence functional, which provides an exact formal expression for the time-evolved density matrix of the quantum dots, and which forms the basis for our further theoretical developments. In Sec.~\ref{sec:largeVs}, we consider the situation where the applied voltages are large, and the electric currents can be evaluated analytically. In Sec.~\ref{sec:heom}, we describe the HEOM technique that we use to calculate the transport properties of the device with finite voltage or temperature differences. In Sec.~\ref{sec:finiteVs}, we then present results for the electric current with finite voltages, which is relevant for the recent experiments. In Sec.~\ref{sec:thermoelec}, we furthermore consider a thermoelectric effect, where the electric current is driven by a temperature difference between the leads, rather than a voltage bias. Finally, in Sec.~\ref{sec:conclus}, we present our conclusions together with an outlook on possible developments for future work. Technical details of the numerics are provided in the Appendix.

\section{Cooper pair splitter}  
\label{sec:CPSmodel}

The Cooper pair splitter in Fig.~\ref{fig1}(a) consists of a superconductor connected to two quantum dots, which are each tunnel coupled to separate normal-state electrodes. \revision{We consider a conventional superconductor with singlet pairing, but note that superconductors with triplet pairing could also be treated. The level spacings of the quantum dots are assumed  to be so large that only a single orbital level of each quantum dot participates in the charge transport. Additional levels could, however, be included in the model below. With a large superconducting gap, the superconductor can be integrated out, and the quantum dots can be described by the  effective Hamiltonian}
\begin{equation}
\label{eq:Heff}
\begin{split}
\hat{H}_{D}&=\sum_{\ell\sigma}\varepsilon_\ell\hat{c}_{\ell\sigma}^\dagger \hat{c}_{\ell\sigma}^{\phantom\dagger} + \sum_{\ell}U  \hat n_{\ell \uparrow} \hat n_{\ell \downarrow}\\
&-\sum_\sigma \kappa \left(\hat{c}_{L\sigma}^\dagger \hat{c}_{R\sigma}^{\phantom\dagger}+
\hat{c}_{R\sigma}^{\dagger}\hat{c}_{L\sigma}^{\phantom{\dagger}}\right)\\
&-\sum_\ell\gamma_\ell \left(\hat{c}_{\ell\uparrow}^\dagger \hat{c}_{\ell\downarrow}^{\dagger}    +\hat{c}_{\ell\downarrow} \hat{c}_{\ell\uparrow}\right)- \gamma\left(\hat d_S^\dagger+\hat d_S\right),
\end{split}
\end{equation}
where the influence of the superconductor enters in the second and third lines~\cite{Sauret2004,Sauret2005,Governale2008,Futterer2009,Meng2009,Eldridge2010,Hiltscher2011,Zonda2023,Walldorf2020,Brange:2021,Tam2021,Brange:2024}. \revision{This approximation assumes that the superconducting gap is the largest energy scale relevant for the low-energy dynamics, including the tunnel couplings, interaction energies, the temperature, and the voltage bias, such that quasiparticle excitations in the superconductor can be neglected and the superconducting proximity effect can be represented by effective pairing amplitudes.}

In the first line, the first term describes the single level of each quantum dot,  $\varepsilon_\ell$, which can be tuned by external gate voltages. The creation and annihilation operators for electrons with spin $\sigma=\uparrow,\downarrow$ are denoted by $\hat{c}_{\ell\sigma}^\dagger$ and $ \hat{c}_{\ell\sigma}$, and  $\hat n_{\ell\sigma}=\hat{c}_{\ell\sigma}^\dagger \hat{c}_{\ell\sigma}^{\phantom\dagger}$ are the corresponding occupation numbers with $\ell=L,R$. The second term describes the interactions between electrons on the same quantum dot with strength $U$. The term in the second line describes the transfer of electrons between the quantum dots, either because of direct tunneling between them or because of elastic cotunneling via the superconductor. The total amplitude of these processes is denoted by $\kappa$. 

The terms in the third line correspond to the transfer of Cooper pairs between the superconductor and the quantum dots. The first term describes the process where two electrons from a Cooper pair tunnel into the same quantum dot, which occurs with the amplitude~$\gamma_\ell$. By contrast, the second term describes the splitting of a Cooper pair, where a singlet state is created between the quantum dots as given by the two-particle operator 
\begin{equation}
\hat d_S^\dagger = (\hat{c}_{L\downarrow}^\dagger \hat{c}_{R\uparrow}^\dagger-\hat{c}_{L\uparrow}^\dagger \hat{c}_{R\downarrow}^\dagger)/\sqrt{2}.
\end{equation}
The amplitude of the nonlocal Cooper pair splitting is suppressed exponentially with the distance between the quantum dots over the coherence length of the \revision{superconductor} as
$\gamma\simeq \sqrt{\gamma_L\gamma_R}\exp(-l/\xi)$~\cite{Sauret2004}. To keep the discussion simple, we take $\gamma_\ell=\gamma_L=\gamma_R$ from now on.

Including the leads, the total Hamiltonian becomes
\begin{equation}
    \hat H = \hat H_{D} + \hat H_N + \hat H_{T},
\end{equation}
where the left and right electrodes are described by
\begin{equation}
    \hat H_N = \sum_{\ell k \sigma} \epsilon_{\ell k} \hat c_{\ell k \sigma}^\dagger \hat c_{\ell k \sigma}^{\phantom\dagger},
\end{equation}
and the transfer of electrons between the quantum dots and the leads is captured by the tunneling Hamiltonian
\begin{equation}
    \hat H_{T} = \sum_{\ell k \sigma} (t_{\ell k} \hat c_{\ell k \sigma}^\dagger \hat c_{\ell \sigma}^{\phantom\dagger} + t_{\ell k}^*  \hat c_{\ell \sigma}^\dagger\hat c_{\ell k \sigma}^{\phantom\dagger}).
\end{equation}
Below, we investigate the charge transport in the device for different voltage and temperature differences.

\section{Fermionic influence functional}
\label{sec:fif}

To describe the transport properties of the Cooper pair splitter, we consider the reduced density matrix of the quantum dots after tracing out the leads. 
Since the electrons in the leads are noninteracting and initially in a thermal state, the  density matrix can be expressed as~\cite{PhysRevB.105.035121}
\begin{equation}
    \hat \rho(t) = \hat{T}\exp [\mathcal{F}(t) ]  \hat \rho(0)
\label{eq:rho_eom}
\end{equation}
in terms of the fermionic influence superoperator 
\begin{equation}
        \mathcal{F}(t) = \int_0^t dt_2 \int_0^{t_2} dt_1 \mathcal{W}(t_2, t_1)
\label{eq:influence}
\end{equation}
and the time-ordering  operator $\hat{T}$. Here, we have defined
\begin{equation}
\begin{split}
\mathcal{W}(t_2,t_1)=\sum_{l\sigma}&\mathcal{A}_{l\sigma}^{+}(t_2)\mathcal{B}_{l\sigma}^+(t_2,t_1)\\
+&\mathcal{A}_{l\sigma}^{-}(t_2)\mathcal{B}_{l\sigma}^-(t_2,t_1) 
\end{split}
\label{eq:W}
\end{equation}
with the superoperators acting on density matrices as
\begin{equation}
\begin{split}
\mathcal{A}_{\ell\sigma}^{+}(t_2)\hat{\rho} = \hat{c}_{\ell\sigma}(t_2)\hat{\rho} - \mathcal{P}[\hat{\rho}\hat{c}_{\ell\sigma}(t_2)],\\
\mathcal{A}_{\ell\sigma}^{-}(t_2)\hat{\rho} = \hat{c}^{\dagger}_{\ell\sigma}(t_2)\hat{\rho} - \mathcal{P}[\hat{\rho}\hat{c}^{\dagger}_{\ell\sigma}(t_2)]
\label{eq:A}
\end{split}
\end{equation}
and
\begin{equation}
\begin{split}
\mathcal{B}^{+}_{\ell\sigma}(t_2,t_1)\hat{\rho} = &- C_{\ell\sigma}^{+}(t_2, t_1)\hat{c}^{\dagger}_{\ell\sigma}(t_1)\hat{\rho} \\&- C_{\ell\sigma}^{-}(t_1, t_2)\mathcal{P}[\hat{\rho}\,\hat{c}^{\dagger}_{\ell\sigma}(t_1)],\\ \label{eq:B}
\mathcal{B}^{-}_{\ell\sigma}(t_2,t_1)\hat{\rho} = &- C_{\ell\sigma}^{-}(t_2, t_1)\hat{c}_{\ell\sigma}(t_1)\hat{\rho} \\&- C_{\ell\sigma}^{+}(t_1, t_2)\mathcal{P}[\hat{\rho}\,\hat{c}_{\ell\sigma}(t_1)].
\end{split}
\end{equation}
Importantly, for fermionic systems, we have to account for the parity of the many-body state using the operator
\begin{equation}
\hat{P}=\exp[i\pi\sum_{\ell\sigma}\hat{c}^{\dagger}_{\ell\sigma}\hat{c}_{\ell\sigma}],
\end{equation}
which yields plus or minus one for states with an even or an odd number of electrons, and the superoperator
\begin{equation}
\mathcal{P}\hat{\rho}=\hat{P} \hat{\rho} \hat{P},   
\end{equation}
which acts on the density matrix of the quantum dots.

Above, we work in the interaction picture with respect to the uncoupled system and leads, $\hat H_D+\hat H_N$, and we have defined the equilibrium bath correlation functions 
\begin{equation}
\begin{split}
    C_{\ell\sigma}^{+}(t_2, t_1) &= \langle\hat B_{\ell\sigma}^\dagger(t_2) \hat B_{\ell\sigma}(t_1)\rangle,\\
    C_{\ell\sigma}^{-}(t_2, t_1) &= \langle\hat B_{\ell\sigma}(t_2) \hat B^\dagger_{\ell\sigma}(t_1)\rangle,
\end{split}
\label{eq:bath_corr}
\end{equation}
in terms of the operators
\begin{equation}
\hat B^\dagger_{\ell\sigma}(t) = \sum_k (t_{\ell k}/\hbar) \hat c_{\ell k \sigma}^\dagger \exp(i\epsilon_{\ell k} t/\hbar).
\end{equation}
The bath correlation functions depend only on the time difference and are independent of the spin, so we can write them as $C_{\ell\sigma}^{\pm}(t_2, t_1)=C_{\ell}^{\pm}(t_2-t_1)$. We also have
\begin{equation}
    C_{\ell}^{\pm} (t) = \int_{-\infty}^\infty \frac{d\omega}{2\pi }  J_\ell(\omega) f_\ell(\pm(\hbar\omega - \mu_\ell)) \exp(\pm i \omega t), 
    \label{eq:c_pm}
\end{equation}
where $f_\ell$ are the Fermi functions of the leads, and we have defined the spectral functions
\begin{equation}
    J_\ell(\omega)= \frac{2\pi}{\hbar} \sum_k |t_{k\ell}|^2\delta(\hbar\omega-\epsilon_{\ell k}).
\end{equation}
We can now take a time-derivative in Eq.~(\ref{eq:rho_eom}) which gives
\begin{equation}     
\partial_t\hat \rho(t) 
     =   \hat{T} \int_0^{t} dt_1 \mathcal{W}(t,t_1) \hat \rho(t).
\end{equation}
It is then convenient to separate $\mathcal{W}(t,t_1)$ into its constituents, because $\mathcal{A}_{\ell \sigma}(t)$ can  be brought in front of the time-ordering operator, since it only depends on the final time $t$. We then arrive at the central equation
\begin{equation}
\begin{split}
    \partial_t\hat \rho(t)  = \sum_{\ell\sigma}&\mathcal{A}^{+}_{\ell\sigma}(t)\hat{T} \int_0^{t} dt_1 \mathcal{B}^{+}_{\ell\sigma}(t,t_1)\hat{\rho}(t)  \\
+&\mathcal{A}^{-}_{\ell\sigma}(t)\hat{T} \int_0^{t} dt_1 \mathcal{B}^{-}_{\ell\sigma}(t,t_1)\hat{\rho}(t),  \label{eq:if_deriv}
\end{split}
\end{equation}
which forms the basis for our further developments.

\section{Large voltages}
\label{sec:largeVs}

We first consider the situation where large positive or negative voltages are applied to the leads. In that case, we can derive a Markovian Lindblad master equation for the density matrix of the quantum dots, which eventually allows us to evaluate the electric currents analytically. 

To derive the Lindblad master equation, we assume flat spectral 
densities for all relevant energies, such that
\begin{equation}
J_{\ell}(     \omega)=\Gamma_{\ell}.
\label{eq:constJ}
\end{equation}
With a large chemical potential, so that the lead becomes a source of electrons, we can take $f_\ell(\hbar  \omega - \mu_\ell)\simeq 1$ and $f_\ell(\mu_\ell-\hbar  \omega )\simeq 0$ in Eq.~(\ref{eq:c_pm}), and we then have $C_{\ell}^{+} (t) = \Gamma_l \delta(t)$ and $C_{\ell}^{-} (t)=0$.
In Eq.~(\ref{eq:if_deriv}), we then find
\begin{equation}
\int_0^{t} dt_1 \mathcal{B}^{+}_{\ell\sigma}(t,t_1)\hat{\rho}(t)  = - \Gamma_\ell\hat{c}^{\dagger}_{\ell\sigma}(t)\hat{\rho}(t)/2 \end{equation}
and
\begin{equation}
\int_0^{t} dt_1 \mathcal{B}^{-}_{\ell\sigma}(t,t_1)\hat{\rho}(t) =
-\Gamma_\ell\mathcal{P}\left[\hat{\rho}(t)\,\hat{c}_{\ell\sigma}(t)\right]/2,
\end{equation}
using the convention that $\int_0^t dt_1 \delta(t-t_1)=1/2$.

Substituting these terms into Eq.~\eqref{eq:if_deriv} and dropping the time ordering, since everything is evaluated at the final time $t$, we find dissipators of the form
\begin{equation}
     \partial_t\hat \rho = \sum_{\ell}\mathcal{D}^{+}_\ell\hat{\rho}=\sum_{\ell\sigma}\Gamma_\ell( \hat{c}^{\dagger}_{\ell\sigma} \hat{\rho}\hat{c}_{\ell\sigma}^{\phantom\dagger} - \{\hat{\rho},\hat{c}_{\ell\sigma}^{\phantom\dagger}\hat{c}^{\dagger}_{\ell\sigma}\}/2),
\label{eq:dissipator_plus}
\end{equation}
which describe how electrons are emitted into the device from the normal leads. Here, we have omitted the time argument and used the fact that the density matrix has even parity so that odd numbers of operators acting on it, under the parity operator, give $\mathcal{P}[\hat{c}_{\ell\sigma} \hat{\rho}]=-\hat{c}_{\ell\sigma} \hat{\rho} $ and so on.

Following a similar line of arguments, one can analyze the situation where the chemical potentials are large and negative, so that the leads function as drains for electrons. In that case, the dissipators become
\begin{equation}
     \partial_t\hat \rho = \sum_{\ell}\mathcal{D}^{-}_\ell\hat{\rho}=   \sum_{\ell\sigma} \Gamma_{\ell} (\hat{c}_{\ell\sigma}^{\phantom\dagger}\hat{\rho}\hat{c}^{\dagger}_{\ell\sigma}-  \{\hat{\rho},\hat{c}^{\dagger}_{\ell\sigma}\hat{c}_{\ell\sigma}^{\phantom\dagger}\}/2),
\label{eq:dissipator_minus}
\end{equation}
which one would also find following Ref.~\cite{Gurvitz1996}. Importantly, the dissipators apply for any system Hamiltonian and for arbitrary couplings to the leads. In the following, we use them to describe the Cooper pair splitter with large (positive or negative) voltages applied to the leads.

We first consider the case where the left lead functions as a source of electrons, while the right lead is a drain. We also assume that the Coulomb interactions are so strong that there is no double occupation of the quantum dots. To implement this restriction, we introduce constrained fermions by redefining the creation operators as $\hat c_{\ell\sigma}^{\dagger}\rightarrow \hat c_{\ell\sigma}^{\dagger}(1-\hat{n}_{\ell\bar\sigma})$, where $\bar\sigma$ denotes the opposite spin of $\sigma$, so that double occupation of the quantum dots is excluded below~\cite{Bozkurt2025}. In the Schr\"odinger picture, we then have 
\begin{equation}
\partial_t\hat \rho=\mathcal{L}\hat \rho=\frac{1}{i\hbar}[\hat H_D,\hat \rho]+\mathcal{D}_L^+\hat \rho+\mathcal{D}_R^-\hat \rho.
\end{equation}
From the Lindblad master equation, we can identify 
\begin{equation}
\mathcal{J}^-_R \hat \rho = \Gamma_R \sum_\sigma  \hat c_{R\sigma}^{\phantom\dagger}\hat \rho \hat c_{R\sigma}^\dagger
\end{equation}
as the jump operator that describes the transfer of electrons into lead $R$. The particle current that runs into the right lead can then be written as
\begin{equation}
 I_R  = \text{tr} \{\mathcal{J}^-_R \hat \rho_s\},
\label{eq:draincurrent}
\end{equation}
where $\hat \rho_s$ is the stationary state, which solves $\mathcal{L} \hat \rho_s = 0$. The particle current that runs into the source electrode, by contrast, is negative and reads 
\begin{equation}
I_L= -\text{tr} \{\mathcal{J}^+_L \hat \rho_s\},
\end{equation}
where 
\begin{equation}
\mathcal{J}^+_L \hat \rho = \Gamma_L \sum_\sigma  \hat c_{L\sigma}^\dagger\hat \rho \hat c_{L\sigma}^{\phantom\dagger}
\end{equation}
is the jump operator that describes the transfer of electrons into the left quantum dot. For a symmetric setup with $\Gamma_L=\Gamma_R=\Gamma$, we then find 
\begin{equation} 
    I_R=-I_L = \frac{12 \Gamma \kappa^2}{4 (\varepsilon_L - \varepsilon_R)^2+9 (\hbar\Gamma)^2 + 18 \kappa^2},
    \label{eq:1source1drain}
\end{equation}
as well as $I_S=I_L+I_R=0$ for the superconductor.

In the case where both leads function as drains for electrons on the quantum dots, we have
\begin{equation}\label{eq:vonNeumannSD}
\partial_t\hat \rho=\mathcal{L}\hat \rho=\frac{1}{i\hbar}[\hat H_D,\hat \rho]+\mathcal{D}_L^-\hat \rho+\mathcal{D}_R^-\hat \rho,
\end{equation}
which is the Lindblad master equation that was used in Refs.~\cite{Sauret2004,Sauret2005,Walldorf2018,Walldorf2020,Tam2021,Brange:2021,Brange:2024} to describe a Cooper pair splitter, where electrons enter the quantum dots from the superconductor and leave via the normal-state drains. For a symmetric setup, we then find~\cite{Walldorf2020,Brange:2024}
\begin{equation}
I_L  =  I_R = \frac{2 \Gamma \gamma^2}{(\varepsilon_L + \varepsilon_R)^2+(\hbar\Gamma)^2+4\gamma^2},
\label{eq:2drains}
\end{equation}
together with $I_S=I_L+I_R>0$ for the superconductor.

In Figs.~\ref{fig1}(b) and \ref{fig1}(c), we show the currents as functions of the level positions for the two voltage configurations. In Fig.~\ref{fig1}(b), we consider the situation where the left lead functions as a source of electrons, while the right lead serves as a drain. In that case, the currents have opposite signs, and they are maximal along the diagonal $\varepsilon_L = \varepsilon_R$, where elastic cotunneling is on resonance, and the currents take the maximum absolute value of 
\begin{equation}
I_L^{\mathrm{max}}  = I_R^{\mathrm{max}} = \frac{4 \Gamma \kappa^2}{3(\hbar\Gamma)^2+6\kappa^2}.
\end{equation}

In Fig.~\ref{fig1}(c), we show the current in the leads, when they both function as drain electrodes. The currents have the same sign and are maximal along the diagonal $\varepsilon_L + \varepsilon_R=0$, where Cooper pair splitting is on resonance, and the currents take the maximum value of  
\begin{equation}
I_L^{\mathrm{max}}  =  I_R^{\mathrm{max}} = \frac{2 \Gamma \gamma^2}{(\hbar\Gamma)^2+4\gamma^2}.
\end{equation}
Thus, from measurements of the currents, one can determine the ratios of the amplitude for Cooper pair splitting and the amplitude for elastic cotunneling over the tunneling rate, $\gamma/(\hbar\Gamma)$ and $\kappa/(\hbar\Gamma)$.

The analytic expressions agree well with the recent experiments on Cooper pair splitters. \revision{In particular, they capture the broadening of the current resonances due to the normal-state electrodes, which would not be included in a weak-coupling description at finite voltage biases. On the other hand, since our analytic expressions} assume large voltage biases, they cannot capture the suppression of the currents, when the quantum dot levels are moved outside of the transport window.

\section{Hierarchical equations of motion}
\label{sec:heom}

To treat finite voltage biases, we now employ the HEOM technique~\cite{Tanimura1989,Tanimura1990}. To this end, we assume that the spectral densities of the leads are Lorentzian,
\begin{equation}
    J_\ell(\omega)=  \frac{\Gamma_\ell W^2}{(\hbar  \omega - \mu_\ell)^2 + W^2},\label{jw}
\end{equation}
centered around the chemical potential of the leads, $\mu_\ell$. \revision{This is the type of spectral function that is currently available in QuTiP, but we note that more general expressions are in the process of being implemented.} If the width $W$ is large compared to other energy scales, we recover the constant spectral function in Eq.~(\ref{eq:constJ}). \revision{Unless stated otherwise, we take $W=50\hbar\Gamma$ with $\Gamma_L=\Gamma_R=\Gamma$.}

For the sake of completeness, we now provide the main steps in the  derivation of the HEOM technique, starting from Eq.~(\ref{eq:if_deriv}).  First, we make the ansatz that the bath correlation functions are given by a sum of exponentials,
\begin{equation}
    C_{\ell}^{\pm} (t) \simeq \sum_{m=1}^{N} \alpha_{ m \ell}^{\pm} \exp(-\gamma_{m \ell }^{\pm} t), \label{exp:ansatz}
\end{equation}
where $\alpha_{m \ell }^{\pm}$ and $\gamma_{ m\ell }^{\pm}$ are fitting parameters. Such an exponential series can be obtained  using a Pad\'{e} decomposition~\cite{pade}. The superoperators in Eq.~(\ref{eq:B}) then become 
\begin{equation}
 \mathcal{B}^{\pm}_{\ell\sigma}(t,t_1)=  \sum_{m=1}^{N}  \mathcal{B}^{\pm}_{m\ell\sigma}(t_1)  e^{-\gamma_{m\ell}^{\pm}(t-t_1)},
 \label{eq:exp_B}
\end{equation}
where we have defined 
\begin{equation}
 \mathcal{B}^{+}_{m\ell\sigma}(t)\hat{\rho} = -\alpha^{+}_{m\ell}\hat{c}^{\dagger}_{\ell\sigma}(t)\hat{\rho} - (\alpha^{-}_{m \ell})^{*}\mathcal{P}[\hat{\rho}\,\hat{c}^{\dagger}_{\ell\sigma}(t)]  
\end{equation}
and
\begin{equation}
 \mathcal{B}^{-}_{m\ell\sigma}(t)\hat{\rho}  =-\alpha^{-}_{m\ell}\hat{c}_{\ell\sigma}(t)\hat{\rho} - (\alpha^{+}_{m \ell})^{*}\mathcal{P}[\hat{\rho}\,\hat{c}_{\ell\sigma}(t)].  
\end{equation}
To arrive at these expressions, we have used the relations $C^{\pm}_{\ell }(-t) = (C^{\pm}_{\ell}(t))^*$ and  $\gamma_{m \ell }^{\pm} = (\gamma_{m \ell }^{\mp})^{*}$.

With this decomposition, Eq.~(\ref{eq:if_deriv}) contains $M = 2\times N \times N_{\ell} \times N_s $ terms like the summand in Eq.~\eqref{eq:exp_B} with $N_{\ell}=2$ reservoirs and $N_s=2$ spin labels, which are described by two correlation functions that are decomposed into $N$ distinct exponentials. We can then rewrite the equation of motion for the density matrix as
\begin{equation}
     \partial_t{\hat \rho}(t) = \sum_{j}\mathcal{A}_{j}(t)\hat \rho_{j}(t),
\label{eq:if_deriv:2}
\end{equation}
where the sum runs over the multi-index $j=(\pm)m\ell\sigma$, and we have defined the auxiliary density operators
\begin{equation}
\hat \rho_{j}(t) = \hat{T}\,\mathcal{D}_{j}(t)\hat{\rho}(t)
\end{equation}
in terms of the superoperators
\begin{equation}
\mathcal{D}_{j}(t)=\int_0^{t} dt_1 \mathcal{B}_{j}(t_1)e^{-\gamma_{j}(t-t_1)}.
\end{equation}
We can now systematically take time derivatives of the auxiliary density operators by employing the Leibnitz integral rule and chain rule, which yield
\begin{equation}
\begin{split}
   \partial_t{ \hat\rho}_{j}(t) = & -\gamma_{j}\hat\rho_{j}(t) +\mathcal{B}_{j}(t)\hat{\rho}(t)\\
    &+ \sum_{j_1} \mathcal{A}_{j_1}(t) \hat{T} \,\mathcal{D}_{j_1}(t)\mathcal{D}_{j}(t)\hat\rho(t).
    \label{eq:partialheom}
\end{split}
\end{equation}
We then see that the equation of motion for the density matrix $\hat{\rho}(t)$ is coupled to $M$ auxiliary density operators. The auxiliary density operators themselves obey equations of motion, which couple them to $\hat{\rho}$, as well as to a new set of $M$ auxiliary density operators from the sum in the second line,  and we consequently define 
\begin{equation}
\hat\rho_{j_1j}(t) = \hat{T}\,\mathcal{D}_{j_1}(t)\mathcal{D}_{j}(t)\hat{\rho}(t)
\end{equation}
together with the general auxiliary density operators
\begin{equation}
\hat\rho_{j_n\cdots j_1}(t)= \hat{T}\,  \mathcal{D}_{j_n}(t) \cdots \mathcal{D}_{j_1}(t)\hat{\rho}(t),
\label{eq:ado1}
\end{equation}
which each obey an equation of motion like Eq.~(\ref{eq:partialheom}). We solve these coupled equations using the implementation of the HEOM technique in QuTiP~\cite{Johansson2012,Johansson2013,Lambert2023,Lambert2026} and then calculate the currents in the Cooper pair splitter. Further details of the numerics are provided in the Appendix.

\begin{figure}
    \centering \includegraphics[width=1\columnwidth]{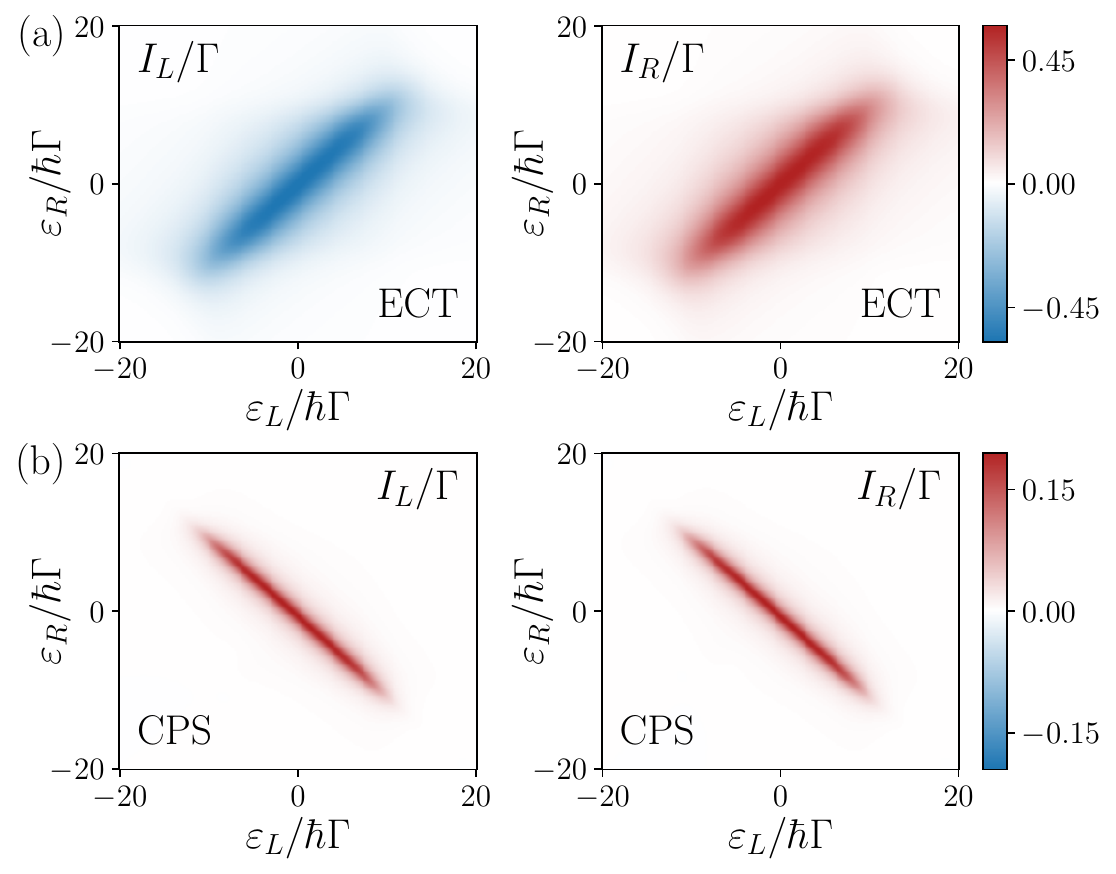}
\caption{Charge transport with finite voltages. (a) A finite voltage is applied between the normal-state electrodes, and elastic cotunneling (ECT) is on resonance, when the two quantum dot levels are aligned. The charge transport is suppressed, if the quantum dot levels are moved outside the voltage window. (b) Here, both normal-state electrodes function as drains, and Cooper pair splitting (CPS) is on resonance, when the two quantum dot levels are oppositely detuned. The charge transport is suppressed, if one of the quantum dot levels is moved below the chemical potential of the leads. The parameters are $\gamma=0.4 \hbar\Gamma$, $\kappa=2 \hbar\Gamma$, $k_B T= 1\hbar \Gamma$, and (a) $\mu_L = -\mu_R = 10 \hbar \Gamma$ and (b) $\mu_L = \mu_R  = -10\hbar \Gamma$.}
    \label{fig2}
\end{figure}

\section{Finite voltages}
\label{sec:finiteVs}
We now turn to calculations of the charge transport with finite voltages and temperatures. First, we consider strong Coulomb interactions, such that double occupation of the quantum dots is excluded. To this end, we use the constrained fermionic operators introduced above. Later, we consider finite Coulomb interactions. At first, we also take a large width of the bath spectral densities, so that they are flat on all relevant energy scales. Later, we consider the effects of a finite width. \revision{We note that Cooper pair splitters have been investigated in the Kondo regime~\cite{Wrzesniewski2017}, which can also be treated with HEOM~\cite{Xu2017}. Here, however, we do not consider such low temperatures that the Kondo effect becomes relevant.}

In Fig.~\ref{fig2}, we show calculations of the electric current in the Cooper pair splitter using HEOM. We use the same parameters as in Fig.~\ref{fig1}; however, we can now include a finite voltage difference and a finite temperature. As a result, the transport is no longer described solely by the resonance conditions of elastic cotunneling and Cooper pair splitting, but is also constrained by the transport windows defined by the chemical potentials of the leads.

Figure \ref{fig2}(a) shows the situation where the left lead functions as a source of electrons and the right lead acts as a drain. The charge transport is then dominated by elastic cotunneling, which is resonantly enhanced when the two quantum dot levels are aligned. However, unlike the expression in Eq.~(\ref{eq:1source1drain}), the resonance is confined to a range of level positions. When either quantum dot level is moved outside the transport window set by the chemical potentials, the current decreases rapidly. The temperature broadens the boundaries of the transport window, leading to a smooth crossover between conducting and nonconducting regions rather than a sharp cutoff. 

In Fig.~\ref{fig2}(b), both leads function as drains, and electrons enter the device from the superconductor through Cooper pair splitting. The current is maximal along the antidiagonal, where Cooper pair splitting is resonant. In contrast to the large-bias result in Eq.~(\ref{eq:2drains}), the finite chemical potentials of the drains strongly influence the current away from resonance. In particular, transport is suppressed when one of the quantum dot levels is moved below the chemical potential of the leads. In that case, the corresponding quantum dot becomes occupied, blocking the transfer of additional electrons into the drain. As a consequence, the current is concentrated within a finite region around the resonance condition, while the surrounding areas exhibit only weak transport.

\begin{figure}
   \centering
\includegraphics[width=1\columnwidth]{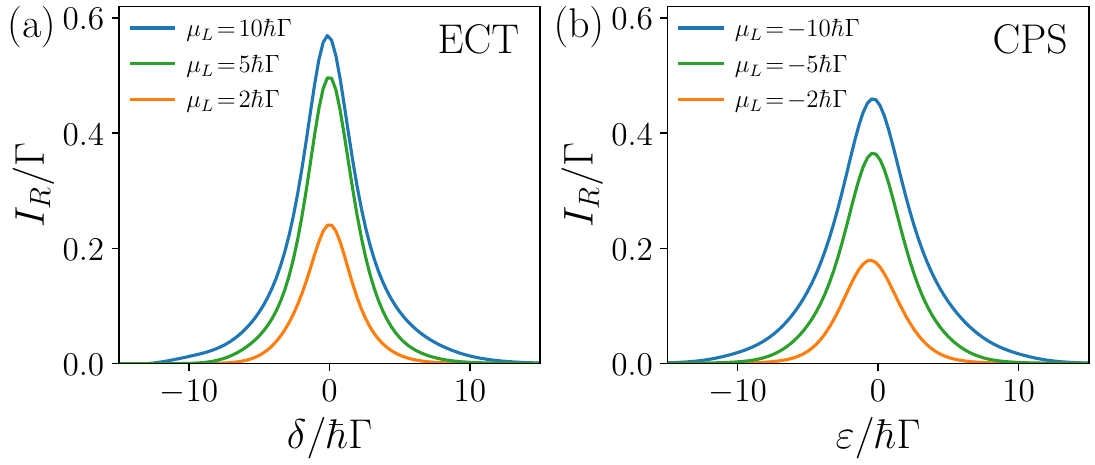}
\caption{Finite voltages. (a) Particle current in the right lead for different voltages across the device with $\mu_L=-\mu_R$, where the transport is dominated by elastic cotunneling (ECT). (b) Particle current in the right lead for different values of the chemical potential in both leads, $\mu_L=\mu_R$, and the transport is dominated by Cooper pair splitting (CPS).  The other parameters are $\gamma = 2 \hbar \Gamma, \kappa = 3 \hbar \Gamma$, and $k_BT = 1 \hbar \Gamma$.}
\label{fig3}
\end{figure}

Comparing Figs.~\ref{fig1} and~\ref{fig2}, we see that the large-bias expressions correctly capture the location of the  resonances, but they overestimate the extent of the conducting regions. The HEOM calculations show how a finite bias  and temperature  modify the transport, which is important for a quantitative comparison with experiments. Indeed, the numerical results in Fig.~\ref{fig2} agree well with the measurements from Ref.~\cite{Wang2022}. Moreover, we can extract realistic parameter values corresponding to our calculations: The maximum values in Fig.~\ref{fig2}(a) correspond to electric  currents of  about $I=0.4 \Gamma e$, while the measured currents were about~$I=0.4$~nA. The tunnel coupling must then be about $\Gamma = 6$ GHz (or $\hbar\Gamma =4$~$\mu$eV) with a temperature of $T=50$ mK for $k_BT=1\hbar\Gamma$.

\begin{figure}
    \centering
    \includegraphics[width=1\columnwidth]{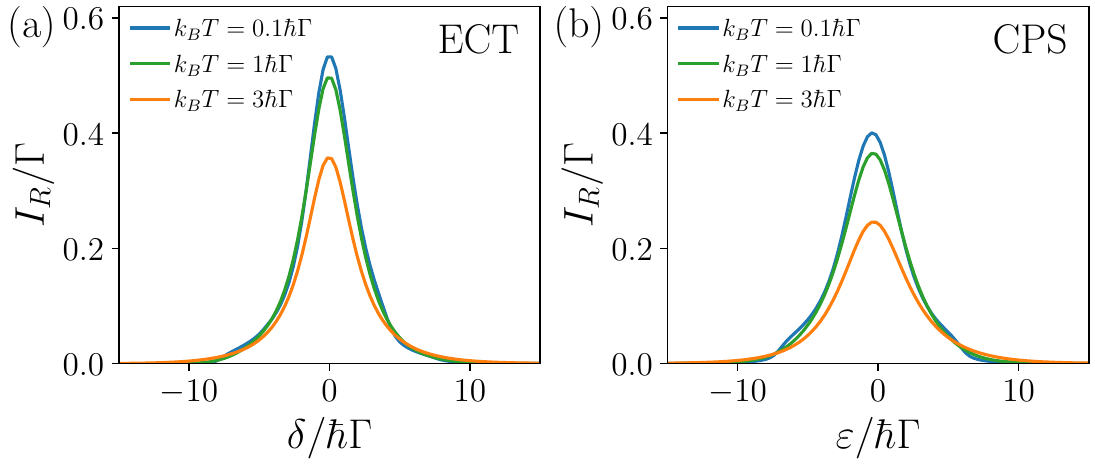}\caption{Finite temperatures. (a) Particle current in the right lead for different temperatures of the leads and opposite chemical potentials, $\mu_L = -\mu_R = 5\hbar \Gamma$, such that the transport is dominated by elastic cotunneling (ECT). (b)   Particle current in the right lead for different temperatures of the leads and the same chemical potentials, $\mu_L = \mu_R = -5\hbar\Gamma$, such that the transport is dominated by Cooper pair splitting (CPS).   The other parameters are $\gamma = 2 \hbar \Gamma$ and $\kappa = 3 \hbar \Gamma$.}
\label{fig4}
\end{figure}

Next, we investigate the influence of the voltage bias, the temperature, finite interactions, and the width of the bath spectral functions. To this end, it will be useful to introduce the detuning of the quantum dot levels,
\begin{equation}
\delta = (\varepsilon_L - \varepsilon_R)/2,
\end{equation}
as well as half of their sum,
\begin{equation}
    \varepsilon = (\varepsilon_L + \varepsilon_R)/2.
\end{equation}

Figure \ref{fig3} illustrates how the finite voltage bias affects transport through the Cooper pair splitter. In Fig.~\ref{fig3}(a), we consider elastic cotunneling and show the current as a function of the detuning for several values of the applied bias. The current exhibits a resonance centered at zero detuning, where elastic cotunneling is most efficient. As the bias voltage is reduced, the resonance peak decreases in height and becomes narrower. This behavior reflects the shrinking transport window available for electrons tunneling between the leads. While the resonance condition itself is unaffected by the bias, a smaller voltage limits the range of energies that contribute to transport.

A similar trend is observed for Cooper pair splitting in Fig.~\ref{fig3}(b). Here, the current is shown as a function of the sum of the quantum dot levels. The current reaches its maximum when the total energy of the quantum dot levels vanishes and Cooper pair splitting is resonant. As the magnitude of the drain voltages is reduced, the current decreases and the resonance becomes increasingly confined. These results demonstrate that finite-bias effects primarily limit the available transport window, while leaving the underlying resonance conditions unchanged.

The influence of temperature is shown in Fig.~\ref{fig4}. For elastic cotunneling in Fig.~\ref{fig4}(a), increasing the temperature leads to  broadening of the resonance around zero detuning. Thermal excitations smear the occupation probabilities in the leads, allowing electrons to participate in transport over a wider energy range. As a result, the current extends beyond the low-temperature resonance region, although the peak value is reduced. The temperature dependence therefore reflects the competition between thermal broadening and the coherent resonant tunneling process responsible for elastic cotunneling. 

\begin{figure}
    \centering
    \includegraphics[width=1\columnwidth]{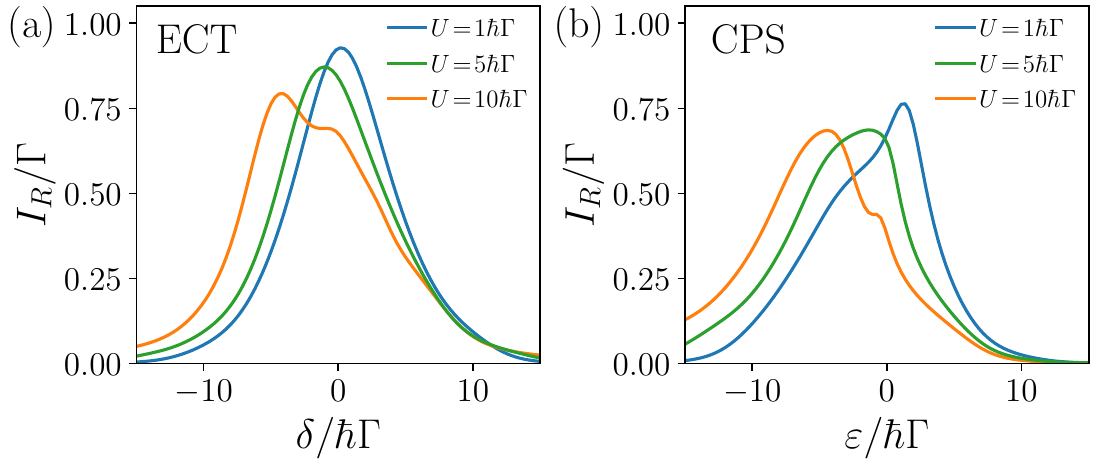}
    \caption{Finite Coulomb interactions.  (a) Particle current in the right lead for different values of the Coulomb interaction $U$ and opposite chemical potentials, $\mu_L = -\mu_R = 10\hbar \Gamma$, such that the transport is dominated by elastic cotunneling (ECT). (b) Particle current in the right lead for different values of the Coulomb interaction $U$ and the same chemical potentials, $\mu_L = \mu_R = -10\hbar \Gamma$, such that the transport is dominated by Cooper pair splitting (CPS).   The other parameters are $\gamma = 2 \hbar \Gamma, \kappa = 3 \hbar \Gamma$, $k_BT=1\hbar \Gamma$, and $\gamma_\ell = \gamma_L = \gamma_R = 4 \hbar\Gamma$ for the amplitude of processes, where a Cooper pair is transferred between the superconductor and one of the quantum dots. }
    \label{fig5}
\end{figure}
For Cooper pair splitting, as shown in Fig.~\ref{fig4}(b), the temperature produces a similar broadening of the resonance. At low temperatures, transport is concentrated close to the resonance condition, whereas higher temperatures allow thermally activated processes to contribute away from resonance. The resulting current profile becomes broader and less sharply peaked. These results indicate that finite temperatures smear the boundaries imposed by the transport window and reduce the contrast between resonant and off-resonant transport.

Figure~\ref{fig5} shows the effect of finite Coulomb interactions on the quantum dots. In contrast to the results discussed above, where double occupation of each quantum dot was excluded, we now allow finite interaction strengths. For elastic cotunneling in Fig.~\ref{fig5}(a), increasing the interaction strength suppresses configurations with double occupancy and gradually drives the system toward the strong-interaction limit considered above. For smaller interaction strengths, additional charge configurations become accessible, modifying both the width and amplitude of the current resonance. As the interaction strength increases, the resonance approaches the behavior expected in the Coulomb-blockade regime.

The corresponding results for Cooper pair splitting are shown in Fig.~\ref{fig5}(b). Cooper pair splitting relies on correlated two-electron processes and is therefore particularly sensitive to the availability of doubly occupied states. For weak interactions, local pair-transfer processes onto the individual quantum dots can compete with nonlocal Cooper pair splitting, modifying both the shape and amplitude of the resonance. As the interaction strength increases, these competing channels are progressively suppressed, and the current evolves toward the strongly interacting limit where Cooper pair splitting dominates. The results highlight the role played by Coulomb interactions in promoting nonlocal Cooper pair splitting.

\revision{We note that the resonance conditions for the local transfer of Cooper pairs into one of the quantum dots read $2\varepsilon_L+U=0$ or $2\varepsilon_R+U=0$ for the left or right quantum dot, since the doubly occupied states now have to be aligned with the chemical potential of the superconductor. The gate-controlled level positions of the quantum dots thereby provide experimental knobs for controlling the different types of Cooper pair transfers and their relative contributions to the electric currents.}

\begin{figure}
    \centering
    \includegraphics[width=1\columnwidth]{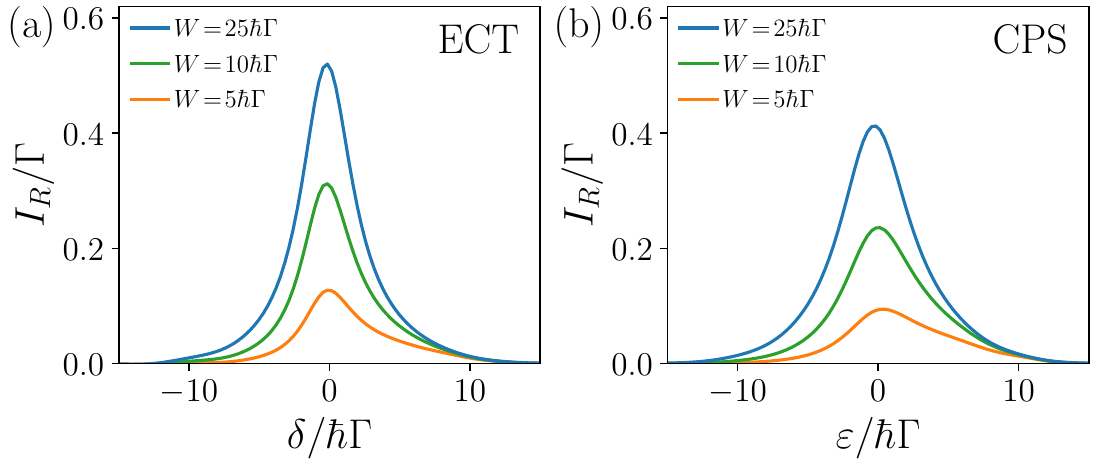}
    \caption{Width of the spectral functions. (a) Particle current in the right lead for different widths of the bath spectral functions $W$ and opposite chemical potentials, $\mu_L = -\mu_R = 10\hbar \Gamma$, such that the transport is dominated by elastic cotunneling (ECT). (b) Particle current in the right lead for different widths of the bath spectral functions $W$ and the same chemical potentials, $\mu_L = \mu_R = -10\hbar \Gamma$, such that the transport is dominated by Cooper pair splitting (CPS). The other parameters are $\gamma = 2 \hbar \Gamma, \kappa = 3 \hbar \Gamma$, and $k_BT= 1\hbar\Gamma$.  }
    \label{fig6}
\end{figure}

Finally, Fig.~\ref{fig6} illustrates the dependence of the current on the width of the Lorentzian spectral density of the leads. In the wide-band limit, the spectral density is effectively flat over all relevant energies. This behavior is seen for the largest width, where the current closely follows the resonance profile expected from the analytical treatment. As the bandwidth is reduced, deviations from the wide-band limit become apparent for both elastic cotunneling and Cooper pair splitting. The finite spectral width reduces the density of available lead states away from the chemical potentials and thereby suppresses transport through off-resonant energy levels. Consequently, the current peaks become lower and narrower as the width decreases. These results demonstrate how non-Markovian effects associated with structured reservoirs influence the transport characteristics of the Cooper pair splitter. At the same time, the resonance positions remain fixed by the  conditions for elastic cotunneling and Cooper pair splitting, indicating that the principal effect of a finite bandwidth is to modify the effective coupling between the quantum dots and the leads.

\section{Thermoelectric effect}
\label{sec:thermoelec}

While charge transport in Cooper pair splitters is most commonly driven by an applied voltage bias, a temperature difference between the normal-state leads can also induce an electric current. Such thermoelectric effects arise from the interplay of superconducting correlations, energy filtering by the quantum dots, and nonlocal transport processes~\cite{Cao2015,Burset2018,Hussein2019,Kirsanov2019,Arora2025,Arrachea2025}, and they have recently been observed experimentally in graphene-based Cooper pair splitters~\cite{Tan2021}. In contrast to conventional thermoelectric transport, the thermoelectric response of a Cooper pair splitter involves both elastic cotunneling and nonlocal Cooper pair splitting processes. The relative importance of these mechanisms depends sensitively on the alignment of the quantum dot levels and the temperature gradient across the device. In the following, we use HEOM to investigate thermoelectric effects in the Cooper pair splitter and analyze how the resulting currents depend on the quantum dot levels and the temperatures of the leads.

\begin{figure}
    \centering
    \includegraphics[width=\columnwidth]{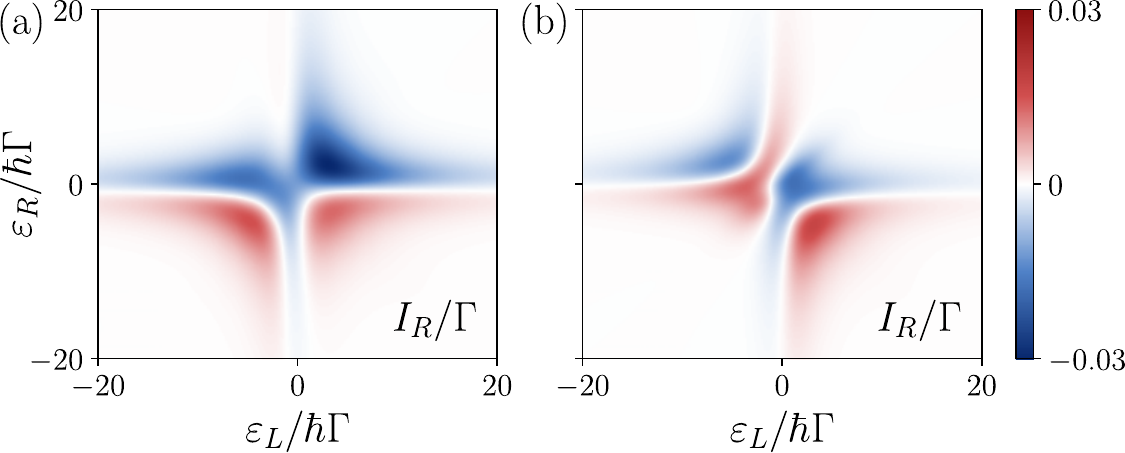}
    \caption{Thermoelectric effect. (a) Particle current induced by a temperature difference between the two normal-state leads with $k_B T_L= 1\hbar\Gamma$ and $k_B T_R= 0.5\hbar\Gamma$, but no applied voltage. The other parameters are $\gamma=3\hbar\Gamma$ and $\kappa=3\hbar\Gamma$.  (b) Similar results for $\gamma=3\hbar\Gamma$ and $\kappa=1\hbar\Gamma$. We show the particle current running \emph{into} the Cooper pair splitter from the right electrode to be consistent with the convention of Ref.~\cite{Burset2018}. }
    \label{fig7}
\end{figure}

\begin{figure}
	\centering
	\includegraphics[width=\columnwidth]{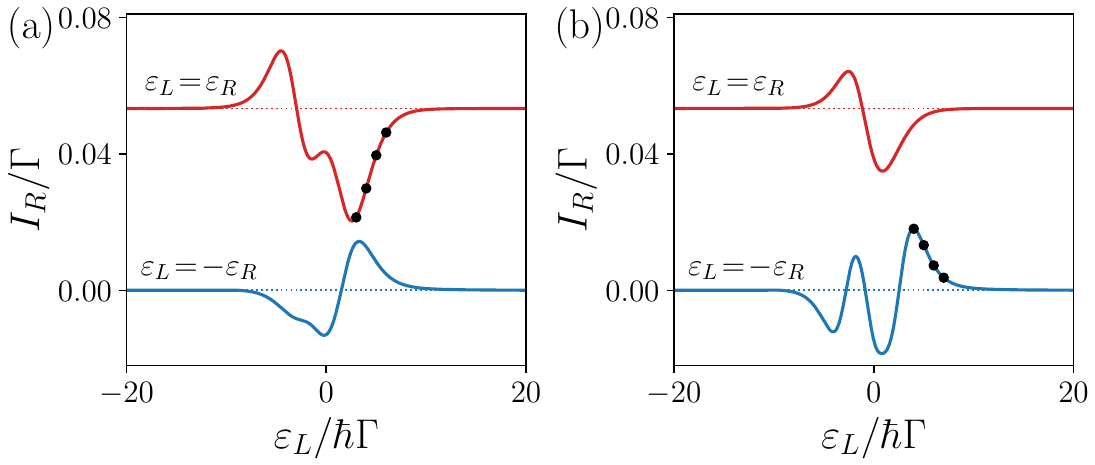}
	\caption{Thermoelectric currents. (a) Thermally induced particle currents along the lines $\varepsilon_L=\varepsilon_R$ and $\varepsilon_L=-\varepsilon_R$ in \revision{Fig.~\ref{fig7}(a)}. (b) Thermally induced particle currents along the same lines in \revision{Fig.~\ref{fig7}(b)}. The curves have been displaced vertically for the sake of clarity. The dots indicate the points for which we calculate the stopping voltage in Fig.~\ref{fig9}.}
	\label{fig8}
\end{figure}

\begin{figure}
	\centering
	\includegraphics[width=\columnwidth]{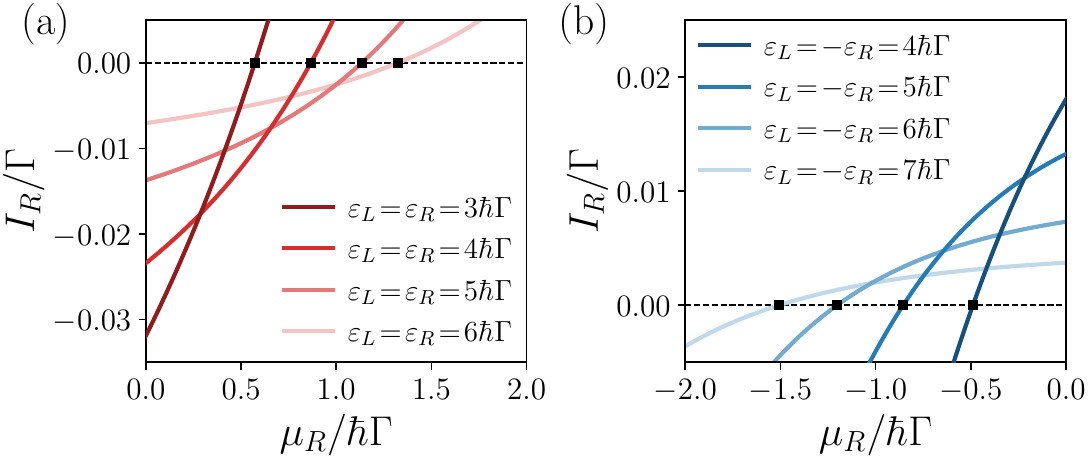}
	\caption{Stopping voltage. (a) Particle current in the right lead as a function of the chemical potential in the right lead. The current vanishes once the stopping voltage (shown with 
		squares) is reached. The curves correspond to the dots in Fig.~\ref{fig8}(a). (b) Similar results for the four dots in Fig.~\ref{fig8}(b).}
	\label{fig9}
\end{figure}

To illustrate these ideas, we consider in Fig.~\ref{fig7} the thermoelectric current generated by a temperature difference between the normal-state leads without an applied voltage. Following the convention of Ref.~\onlinecite{Burset2018}, we show the particle current flowing from the right lead into the Cooper pair splitter as a function of the quantum dot levels.  The resulting current maps are shown in Fig.~\ref{fig7}, selected cuts through these maps in Fig.~\ref{fig8}, and the corresponding calculations of the stopping voltage in Fig.~\ref{fig9}.

A finite thermoelectric current requires the breaking of the energy-inversion symmetry of the device. If electron- and holelike transport processes contribute equally, the thermally induced particle and hole currents compensate each other, resulting in zero net charge current. By tuning the quantum dot levels away from the energy-inversion symmetric points, this cancellation is lifted, allowing the thermal bias to generate a finite electrical current whose sign reflects the dominant charge carriers.

The thermoelectric response is closely tied to the underlying charge stability diagram of the double quantum dot. The largest currents occur along the boundaries separating different charge stability regions, where two or more charge states become degenerate and charge fluctuations are enhanced. By contrast, transport is strongly suppressed inside the Coulomb-blockaded regions, where a single charge configuration is energetically favored. 

The overall structure of the current maps agrees well with the weak-coupling results in Ref.~\onlinecite{Burset2018}. Here, however, we can treat stronger couplings to the leads, where the broadening of the quantum dot levels becomes significant. Consequently, the resonances predicted by perturbative master-equation approaches evolve into broadened features that persist over a wider range of gate voltages. These effects are naturally captured by the HEOM approach and go beyond weak-coupling treatments.

Finally, we turn to the stopping voltage, which is the voltage that  compensates the thermally induced charge current. At this operating point, the electrical and thermal driving forces balance, resulting in zero net current. In contrast to the current itself, the stopping voltage is determined by the balance of competing transport processes and is therefore comparatively insensitive to the  magnitude of the tunnel couplings, making it a robust quantity for comparison between theory and experiment. 

In Fig.~\ref{fig9}, we show calculations of the stopping voltage using HEOM. In Fig.~\ref{fig9}(a), the stopping voltage is positive and increases with the sum of the level positions, while the corresponding values in Fig.~\ref{fig9}(b) are negative and become increasingly large in magnitude as the sum of the level positions is decreased. This behavior reflects the change in the dominant transport processes as the  level positions are tuned relative to the Fermi energy. The stopping voltage therefore provides a direct measure of the particle-hole asymmetry responsible for the thermoelectric current. In particular, the opposite signs observed for the two sets of level configurations are consistent with the reversal of the dominant charge carriers when the levels are tuned from electronlike to holelike transport. The results show that the thermoelectric response can be compensated by a voltage bias, whose magnitude and sign are controlled by the tunable levels. 

\section{Conclusions and outlook}
\label{sec:conclus}

We have investigated charge transport in Cooper pair splitters using hierarchical equations of motion (HEOM), which allow us to treat finite voltage and temperature biases beyond the weak-coupling and Markovian limits. With large voltages, our approach yields analytical results based on Markovian Lindblad master equations, which provide a useful benchmark for the numerical calculations. Away from the large-bias limit, HEOM captures the combined effects of finite transport windows, temperature broadening, and interactions. We have shown that these effects strongly influence the current resonances associated with elastic cotunneling and Cooper pair splitting, including a suppression of transport when the quantum dot levels are moved outside the transport window. We have also investigated the influence of finite Coulomb interactions and nonflat spectral densities of the leads, demonstrating the flexibility of the HEOM framework for describing realistic devices. Finally, we considered thermoelectric effects in Cooper pair splitters, which can arise when the energy-inversion symmetry of the device is broken. Our calculations reproduce key features observed in recent experiments on Cooper pair splitters, including thermoelectric transport induced by temperature differences.

The present work opens several directions for future research. It would be interesting to use HEOM to investigate spin-resolved transport and entanglement generation in devices supporting singlet and triplet Cooper-pair splitting, as realized in recent experiments. Furthermore, HEOM could be used to investigate time-dependent driving protocols, including gate-voltage modulation and pumping schemes, where non-Markovian effects may play an important role. Another promising direction is the application of the method to coupled Cooper pair splitters and short Kitaev chains, where interactions, finite temperatures, and nonequilibrium conditions could influence the emergence of topological superconductivity and Majorana-like bound states. More broadly, our work establishes HEOM as a powerful and versatile tool for investigating nonequilibrium transport phenomena in interacting superconducting nanostructures.

\begin{acknowledgments}
We thank P.~Burset and B.~Muralidharan for useful discussions and acknowledge support from the Nokia Industrial Doctoral School in Quantum Technology and the computational resources provided by the Aalto Science-IT project. This article is based upon work from COST Action ``Many-body Open Quantum Systems" (QOpen) CA24109, supported by COST (European Cooperation in Science and Technology). F.N.~is supported in part by the Japan Science and Technology Agency (JST) [via the CREST Quantum Frontiers program Grant No.~JPMJCR24I2, the Quantum Leap Flagship Program (Q-LEAP), and the Moonshot R\&D Grant Number JPMJMS2061] and by the Office of Naval Research (ONR) Global (via Grant No.~N62909-23-1-2074).  N.L.~is  supported by MEXT KAKENHI (Grant No.~JP24H00816 and Grant No.~JP24H00820).
\end{acknowledgments}

\section*{DATA AVAILABILITY}

The data and codes that support the findings of this article are openly
available~\cite{GitHub:2026}.

\appendix
\section{Numerical details}
\label{app:num_detail}

For the numerics, it is useful to write the auxiliary density operators as
$\hat\rho^{(\bar{n})}$, using the index of integers $\bar{n}=[n_{M},\cdots , n_1]$, where $n_k$ counts the unique instances of $j_k$ in $\mathcal{D}_{j_k}(t)$ in Eq.~(\ref{eq:ado1}), which can only be 0 or 1, since $\mathcal{D}_{j_k}^2(t)=0$. Upon returning to the Schr\" odinger  picture, we can write Eq.~(\ref{eq:partialheom}) as
\begin{equation}
\begin{split}
    \partial_t{\hat \rho}^{(\bar{n})} =&\frac{1}{i\hbar}[\hat H_D,\hat \rho^{(\bar{n})}]
    - \sum_{k=1}^{M} n_k \gamma_{j_k}\hat \rho^{(\bar{n})}  \\
    &+ \sum_{k=1}^{M} (-1)^{f(\bar{n},k)} \mathcal{B}_{j_k} \hat \rho^{[n_k-1]}  \\ & + \sum_{k=1}^{M} (-1)^{f(\bar{n},k)}\mathcal{A}_{j_k}  \hat \rho^{[n_k + 1]}, 
    \label{eq:heom_final}
\end{split}
\end{equation}
where $f(\bar{n},k)=\sum_{l={k+1}}^{M} n_l$ counts the number of nonzero indices to the left of $n_k$ in $\bar{n}$, and the notation $\rho^{[n_k\pm 1]}$ indicates the auxiliary density operator obtained by increasing or decreasing $n_k$ by 1 in $\bar{n}$. The parity operators in Eqs.~(\ref{eq:A}) and (\ref{eq:B}) can now be evaluated, yielding
\begin{equation}
\begin{split}
    \mathcal{A}_{+m\ell \sigma} \hat \rho^{(\bar{n})} &= \hat c_{\ell \sigma} \hat \rho^{(n)} + (-1)^{f(\bar{n},0)} \hat \rho^{(n)} \hat c_{\ell \sigma}, \\
    \mathcal{A}_{-m\ell \sigma} \hat \rho^{(\bar{n})} &= \hat c^{\dagger}_{\ell \sigma} \hat \rho^{(n)} + (-1)^{f(\bar{n},0)} \hat \rho^{(n)} \hat c^{\dagger}_{\ell \sigma} 
    \end{split}
\end{equation}
and 
\begin{equation}
\begin{split}
    \mathcal{B}_{+m\ell \sigma}  \hat\rho^{(n)} &= -\alpha^{+}_{m \ell} c^{\dagger}_{\ell\sigma} \hat \rho^{(n)} + (-1)^{f(\bar{n},0)}(\alpha^-_{m \ell})^* \hat \rho^{(n)}  \hat c^{\dagger}_{\ell\sigma}, \\
    \mathcal{B}_{-m\ell \sigma}  \hat\rho^{(n)} &= -\alpha^-_{m \ell} c_{\ell\sigma} \hat \rho^{(n)} + (-1)^{f(\bar{n},0)} (\alpha^{+}_{m \ell})^*\hat \rho^{(n)}  \hat c_{\ell\sigma}.
\end{split}
\end{equation}

To find the electric currents, we define the particle current operator for lead $\ell$ as the rate of change of particles in the lead, $\hat{I}_{\ell}= [\hat N_{\ell},\hat H]/(i \hbar)$, with occupation number operator $\hat{N}_{\ell}=\sum_{k\sigma} \hat c_{\ell k\sigma }^{\dagger}\hat c_{\ell k \sigma}$. We then obtain
\begin{equation}
     \hat{I}_{\ell} =\frac{1}{i\hbar} \sum_{k\sigma}\left(t_{\ell k}\hat c^{\dagger}_{\ell k\sigma }\hat c_{\ell  \sigma}  -t_{\ell k}^*\hat c_{\ell\sigma}^{\dagger}\hat c_{\ell k \sigma} \right).
\end{equation}
To evaluate the average current within the HEOM technique, we proceed as in Ref.~\cite{Jin2008} and  write the the equation of motion for the reduced density matrix in terms of the full density matrix as 
\begin{equation} 
    \partial_t \hat \rho =  \frac{1}{i\hbar}\sum_{\ell k\sigma}\mathrm{tr}_N\left[ t_{\ell k} \hat c_{\ell k \sigma }^{\dagger}\hat{c}_{\ell \sigma}+ t_{\ell k}^*\hat c_{\ell \sigma}^{\dagger} \hat c_{\ell k \sigma }, \hat \rho_{\mathrm{tot}}\right].
\end{equation}
From Eq.~(\ref{eq:if_deriv:2}), we also have
\begin{equation}
     \partial_t{\hat \rho} = \sum_{m\ell\sigma} ([\hat{c}_{\ell\sigma},\hat{\rho}_{+m\ell \sigma}]
+[\hat{c}^{\dagger}_{\ell\sigma},\hat{\rho}_{-m\ell \sigma}]). \label{eq:if_deriv:3}
\end{equation}
By tracing over the quantum dots and comparing the two expressions, we can match terms like
\begin{equation}
-\frac{1}{i\hbar}\sum_k\mathrm{tr}_D[\hat c_{\ell \sigma} \mathrm{tr}_N[ t_{\ell k}\hat c_{\ell k \sigma }^{\dagger}\hat \rho_{\mathrm{tot}}]] = \sum_m \mathrm{tr}_D[\hat c_{\ell\sigma} \hat \rho_{+m\ell \sigma}]
\end{equation}
and
\begin{equation}
\frac{1}{i\hbar}\sum_k\mathrm{tr}_D[\hat c_{\ell \sigma}^{\dagger} \mathrm{tr}_N\left[ t_{\ell k}^*\hat c_{\ell k \sigma}\hat \rho_{\mathrm{tot}} \right]] = \sum_m \mathrm{tr}_D[\hat c_{\ell\sigma}^{\dagger} \hat \rho_{-m\ell \sigma}].
\end{equation}
Finally, for the average current, we then get
\begin{equation}
I_\ell =\mathrm{tr}[\hat{I}_{\ell}\hat \rho_{\mathrm{tot}}]= \sum_{m\sigma} \mathrm{tr}_D[\hat c_{\ell\sigma} \hat \rho_{+m\ell \sigma}-\hat c_{\ell\sigma}^{\dagger} \hat \rho_{-m\ell \sigma}],
\end{equation}
which can be found from the auxiliary density matrices.

\revision{For our calculations, we have used a hierarchy depth of 2 together with 18 terms in the Padé expansion. We have also checked that our results have converged upon increasing the number of terms in the Padé expansion.}

%

\end{document}